\documentclass[aps,pre,twocolumn,showpacs,floatfix,superscriptaddress]{revtex4}
\usepackage{amssymb}
\usepackage{amsmath}
\usepackage{graphicx}
\usepackage{latexsym}
\usepackage{amsfonts}
\usepackage{color}

\begin{document}

\title{Symmetry breaking and restoring wave transmission in diode-antidiode
double chains}
\date{\today}
\author{Stefano Lepri}
\email{stefano.lepri@isc.cnr.it}
\affiliation{Consiglio Nazionale delle Ricerche, Istituto dei Sistemi Complessi, via
Madonna del Piano 10, I-50019 Sesto Fiorentino, Italy}
\author{Boris A. Malomed}
\affiliation{Department of Physical Electronics, School of Electrical Engineering,
Faculty of Engineering, Tel Aviv University, Tel Aviv 69978, Israel}

\begin{abstract}
We introduce a system of two parallel-coupled discrete nonlinear Schr\"{o}%
dinger (DNLS) inhomogeneous chains. Each one favors the unidirectional
transmission of incident packets, in the \emph{opposite directions} with
respect to each other. Two different configurations of the diode-antidiode
pair are considered, a ladder and a plaquette. They feature, respectively,
the uniform transverse linear coupling, or the coupling limited to the
central nonlinear segment of the system. In the case of weak linear
coupling, the symmetry breaking is observed (i.e., the pair still features
the diode behavior), while the moderately strong coupling restores the
symmetry, making the transmission effectively bidirectional. In the case of
the ladder, an oscillatory dependence of the transmission on the strength of
the coupling is observed and qualitatively explained.
\end{abstract}

\maketitle

\section{Introduction}

Asymmetric (nonreciprocal) wave propagation induced by nonlinearity is of
great interest as a basic principle for the energy-flow control. It emerges
in several different context. Among the first examples discussed in the
literature is the asymmetric phonon transmission through a nonlinear
interface layer between two very dissimilar crystals \cite{Kosevich1995}. In
the realm of acoustics the possibility of realizing a wave diode has been
demonstrated for nonlinear phononic media \cite{Liang09,Liang2010}. Another
promising context is the propagation of acoustic pulses through granular
systems. In the latter context, experimental studies have demonstrated a
change of the reflectivity of solitary waves from the interface between two
granular media \cite{Nesterenko05}. More recently, demonstration of
rectification of the mechanical-energy transfer at sonic frequencies in a
one-dimensional array of particles has been reported too \cite{Boechler2011}%
. In the field of nonlinear optics, the so-called all-optical diode was
first proposed in Ref.~\cite{Scalora94,Tocci95}, and later on realized
experimentally \cite{Gallo01}. Other diverse implementations of the
unidirectional transmission have been suggested, ranging from left-handed
materials \cite{Feise05}, periodic \cite{Konotop02} and quasiperiodic \cite%
{Biancalana08} photonic crystals, coupled nonlinear cavities \cite%
{Grigoriev2011} to $\mathcal{PT-}$symmetric waveguides \cite%
{Ramezani2010,D'Ambroise2012}. An extension to quantum regimes, in which
few-photon states may also display a diode effect, has been proposed in Ref.
\cite{Roy2010}. A related work for electric transmission lines was reported
in Ref.~\cite{Tao2011}. Interestingly, for nonlinear lattices brought in
contact with thermal baths reciprocity violations are related to the second
principle of thermodynamics \cite{Narayan04}. Recently, a critical
discussion of the topic, including the case of thermal sources, was given in
Ref.~\cite{Maznev2012}.

A reference model to discuss this type of phenomena is the Discrete
Nonlinear Schr\"{o}dinger (DNLS) equation \cite{Eilbeck1985} with embedded
nonlinear elements \cite%
{Molina1993,Gupta1997,Gupta1997a,Bulgakov2011,Brazhnyi2011}. Beyond its
relevance in various physical contexts, ranging from nonlinear optics to the
physics of trapped cold gases \cite{Kevrekidis,Eilbeck2003}, the DNLS
equation has the major advantage of being one of simplest dynamical systems
amenable to a complete theoretical analysis. In the present context,
DNLS-based models are particularly relevant, as they allow one to solve the
scattering problem exactly, without complications of dealing with wave
harmonics \cite{Lepri2011}. In particular, it has been demonstrated in Ref.
\cite{Lepri2011} that the effect of the nonlinear shift of a resonance can
be exploited to design a \textit{diode chain}, capable to transmit waves
with the same amplitude and frequency differently in opposite directions.

The propagation direction admitted by a diode chain is reversed in a chain
which is a specular counterpart of the given one. This obvious fact suggests
us to consider the transmission of waves in double chains, formed by two
parallel diode cores with \emph{opposite orientations}, which are linearly
coupled to each other in the transverse direction. This setting, which may
be called a \textit{diode-antidiode pair} (alias a \textit{diode dipole}),
is the subject of the present work. The transmission of waves in either
direction in such a system obviously implies a spontaneous breaking of the
antisymmetry between the transversely coupled counter-oriented cores. It is
relevant to mention that the spontaneous symmetry breaking in a system of
linearly coupled identical one-dimensional lattices with the onsite cubic
nonlinearity, which admit the transmission of waves in both directions, was
studied in Ref. \cite{Herring2007}. That system was modeled by two linearly
coupled DNLS equations, and the symmetry breaking was considered for
quiescent discrete solitons. The spontaneous (anti)symmetry-breaking effect,
addressed in the present work, is completely different, being dynamical,
rather than static, and realized in terms of incident, transmitted, and
reflected linear waves, rather than solitons. In this respect, this is a new
setting for the study of spontaneous-symmetry-breaking effects, which
manifest themselves in a great variety of nonlinear systems \cite%
{Malomed2013}.

The rest of the paper is organized as follows. Two different models of the
diode-antidiode pair are formulated in Section \ref{sec:mod}. In Section \ref%
{sec:sim} we describe simulations of the wave-packet scattering. In Section %
\ref{sec:break} we reports results for various dynamical regimes, focusing
on the spontaneous symmetry breaking and restoration for the two different
types of the coupled diode-antidiode system. We complete the paper with
concluding remarks in Section \ref{sec:fin}.

\section{Models}

\label{sec:mod}

We consider two linearly-coupled DNLS chains, with onsite dynamical complex
variables, $u_{n}(t)$ and $v_{n}(t)$, where $n$ is the discrete coordinate
in the one-dimensional chains, and the overdot stands for the time
derivative:
\begin{eqnarray}
i\dot{u}_{n} &=&U_{n}u_{n}-u_{n+1}-u_{n-1}+\alpha
_{n}|u_{n}|^{2}u_{n}+\kappa _{n}v_{n},  \notag \\
i\dot{v}_{n} &=&V_{n}v_{n}-v_{n+1}-v_{n-1}+\beta _{n}|v_{n}|^{2}v_{n}+\kappa
_{n}u_{n},  \label{tdnls}
\end{eqnarray}%
The chains are non-uniform, with the real onsite nonlinearity coefficients, $%
\alpha _{n}$ and $\beta _{n}$, as well as the real transverse-coupling
coefficient, $\kappa _{n}$, depending on $n$. In addition, each chains
carries a discrete linear potential, represented by real discrete functions $%
U_{n}$ and $V_{n}$.

\begin{figure}[th]
\begin{center}
\includegraphics[width=0.5\textwidth,clip]{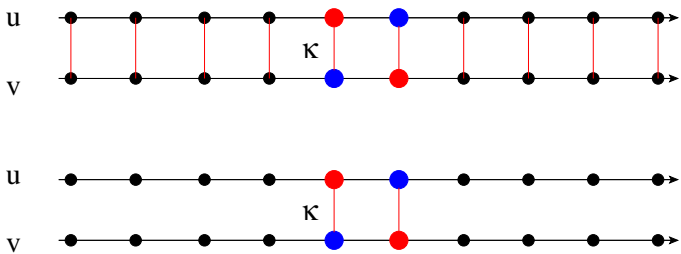}
\end{center}
\caption{(Color online) A sketch of the two models (for $N=2$): the
``ladder" in the upper panel, and the ``plaquette" in the lower one.
In both models, the nonlinearity acts at two central sites of the
chains. } \label{fig:model}
\end{figure}

As sketched in Fig.~\ref{fig:model}, we will consider two versions of the
general model. The first system includes the uniform linear coupling, while
in the other one the coupling acts only at two central sites (with $n=1$ and
$n=2$):
\begin{gather}
\kappa _{n}\equiv \kappa ~~\text{for every $n,$}  \label{plaquette} \\
\kappa _{n}=\kappa \,(\delta _{n,1}+\delta _{n,2}).  \label{ladder}
\end{gather}%
We will refer to these two models as the \textit{ladder} or \textit{plaquette%
} , respectively. In fact, the ladder configuration for static discrete
solitons in the usual bidirectional system was introduced in Ref. \cite%
{Herring2007}, while the plaquette resembles the bidirectional system with
the coupling applied at the single central site \cite{Hadvzievski2010}.

We assume the usual scattering setup which implies that $V_{n}$, $U_{n}$, $%
\alpha _{n}$ and $\beta _{n}$ in Eqs. (\ref{tdnls}) are nonvanishing only in
a finite segment of the chain,
\begin{equation}
1\leq n\leq N.  \label{N}
\end{equation}%
The two semi-infinite portions of the lattice ($n<1$, $n>N$) represent two
uniform linear leads, where waves propagate freely. In the ladder
configuration, the free propagation obeys the following dispersion relation
for wavenumber $k$ and frequency $\omega $:
\begin{equation}
\omega (k)=-2\cos k\pm |\kappa |,  \label{omega}
\end{equation}%
the respective group velocity being%
\begin{equation}
V_{\mathrm{gr}}\equiv d\omega /dk=2\sin k.  \label{V}
\end{equation}

Actually, we consider the dimer, with $N=2$ in Eq. (\ref{N}), i.e.,
\begin{eqnarray}
&&U_{n}=V_{0}\left[ (1+\varepsilon )\delta _{n,1}+(1-\varepsilon )\delta
_{n,2}\right] ,  \label{D1} \\
&&V_{n}=V_{0}\left[ (1-\varepsilon )\delta _{n,1}+(1+\varepsilon )\delta
_{n,2}\right] .  \label{D2} \\
&&\alpha _{n}=\beta _{n}=\delta _{n,1}+\delta _{n,2}.  \label{D3}
\end{eqnarray}%
The system is invariant with respect to the mirror reflection about the
dimer's center, $n\rightarrow N+1-n$, combined with interchange of the
chains $(u_{n},v_{n})\rightarrow (v_{n},u_{n})$. %\[
% (u_n,v_n) \to (v_n,u_n)
%\]

The dimer in the single chain ($\kappa =0$) was considered in Ref. \cite%
{Lepri2011}. Numerical simulations have demonstrated that Gaussian wave
packets impinging on the dimer from the two opposite directions have very
different transmission coefficients, which implies that the single chain
with the embedded dimer plays the role of the diode for the wavepacket
transmission.

\section{Simulations of the wavepacket scattering}

\ \label{sec:sim}

We performed simulations of Eqs. (\ref{tdnls}) for the chains of finite
lengths, i.e. for $|n|\leq M$ (this means that each of the two chains is composed 
of $2M+1$ sites). Open boundary conditions are enforced on both 
chains. Initial condition were
taken as two Gaussian wavepackets, with $n_{0}<0$:
\begin{eqnarray}
&&u_{n}(0)\;=\;\;I(1+\delta )\exp \left[ -\frac{(n-n_{0})^{2}}{w^{2}}+ik_{0}n%
\right] ,  \notag \\
&&v_{n}(0)\;=\;\pm I(1-\delta )\exp \left[ -\frac{(n-n_{0})^{2}}{w^{2}}%
+ik_{0}n\right] ,  \label{gauss}
\end{eqnarray}%
where $\delta $ is a small perturbation (typically we set $\delta =10^{-4}$
henceforth), which seeds symmetry-breaking effects. The symmetries of the
models suggest to consider two types (even and odd) of initial conditions (%
\ref{gauss}), with plus and minus signs, respectively, in the expression for
$v_{n}(0)$. Note that the total initial input power [which is a dynamical
invariant of Eqs. (\ref{tdnls})] is $\sum_{n}\left(
|u_{n}(0)|^{2}+|v_{n}(0)|^{2}\right) \propto w|I|^{2}$ (up to a small
correction $\sim \delta $).

The wavepacket-transmission coefficients in the two coupled chains, produced
by the simulations in the course of temporal interval $0<t<t_{\mathrm{fin}}$%
, are naturally defined as%
\begin{eqnarray}
&&t_{u}=\frac{\sum_{n>N}|u_{n}(t_{\mathrm{fin}})|^{2}}{%
\sum_{n<1}(|u_{n}(0)|^{2}+|v_{n}(0)|^{2})},  \label{tu} \\
&&t_{v}=\frac{\sum_{n>N}|v_{n}(t_{\mathrm{fin}})|^{2}}{%
\sum_{n<1}(|u_{n}(0)|^{2}+|v_{n}(0)|^{2})}.  \label{tv}
\end{eqnarray}%
To minimize the dispersive effects and, thus,
the dependence of the scattering process on the initial position $n_{0}$, we
fix $k_{0}=\pi /2$ in Eqs. (\ref{gauss}).
Moreover, for a given lattice size $M$, $t_{\mathrm{fin}}$ is chosen to be not too
large, so as to avoid hitting the boundary sites by the transmitted and
reflected packets ($M=500$ and $t_{\mathrm{fin}}=250$ are the values used 
in the following). 

\begin{figure}[th]
\begin{center}
\includegraphics[width=0.5\textwidth,clip]{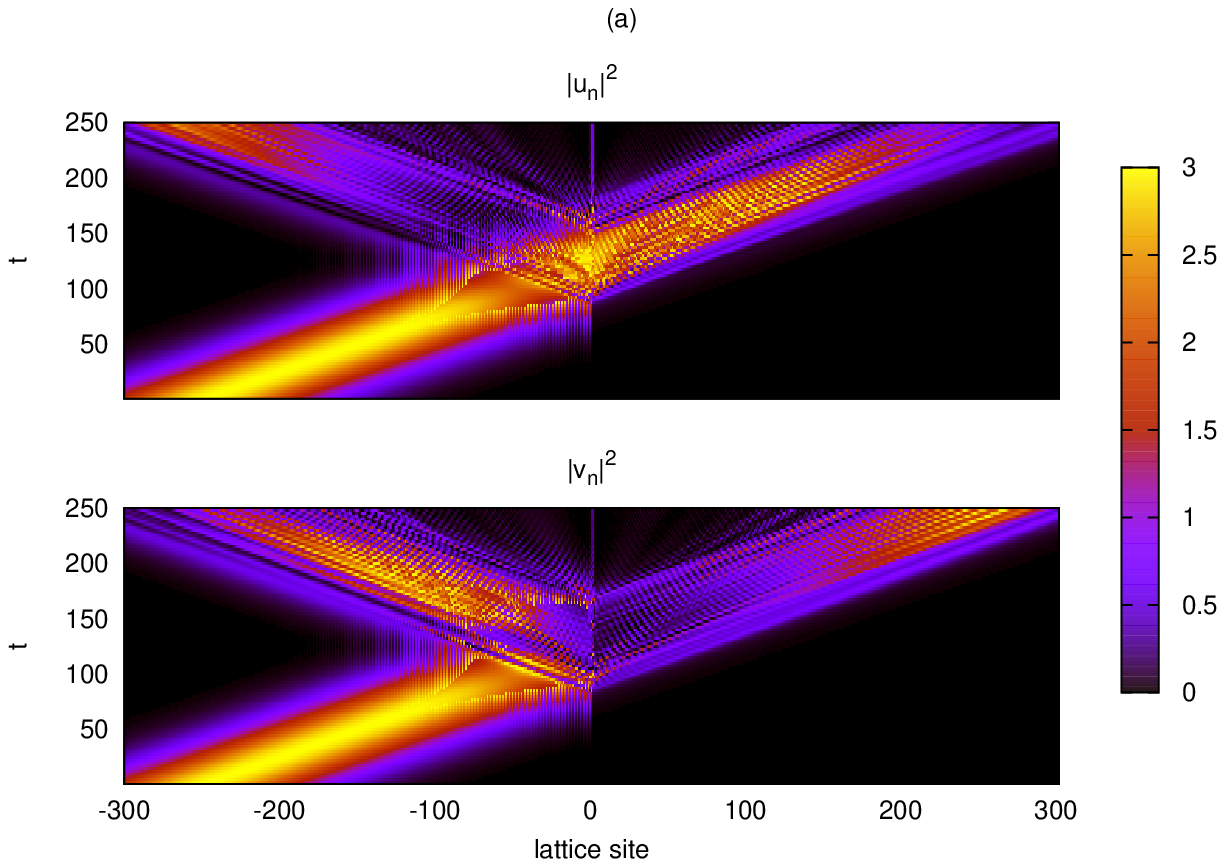}
\par
%\vspace{-1.5cm}
\par
\includegraphics[width=0.5\textwidth,clip]{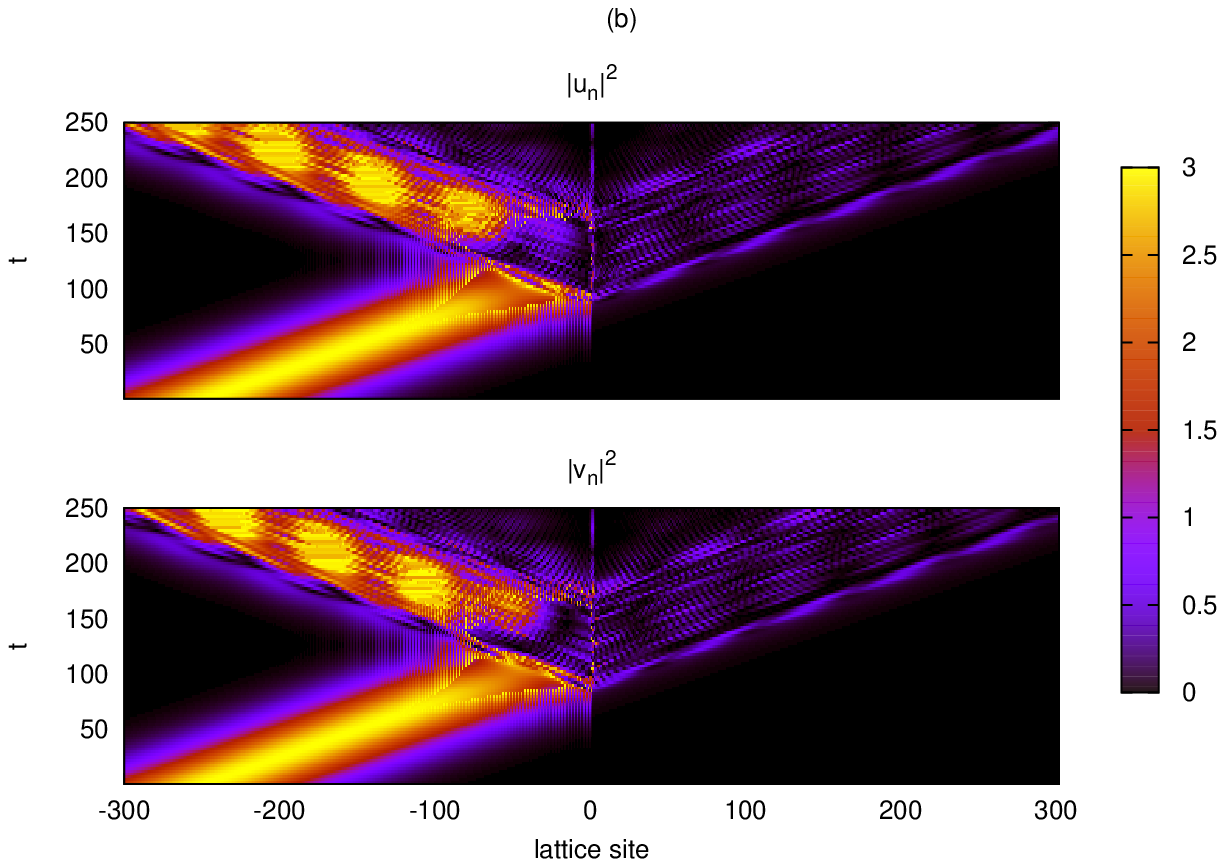}
\end{center}
\caption{(Color online) Numerical simulations of the transmission of
Gaussian wavepackets~(\protect\ref{gauss}) in the ladder system. Parameters
are $V_{0}=-2.5$, $k_{0}=\protect\pi /2$, $\protect\varepsilon =0.05$, $%
M=500 $, $|I|^{2}=3$, $w=100$ and $n_{0}=-250$. Panels (a) and (b) pertain,
respectively, to the weak and moderate couplings, $\protect\kappa =0.01$ and
$\protect\kappa =0.1$.}
\label{f:fig5}
\end{figure}

First, Fig.~\ref{f:fig5}a shows that, for the case of very weakly coupled
chains, with $\kappa =0.01$, the transmission of the incident packet is
strongly different in the two components ($t_{u}=0.34$, $t_{v}=0.15$), as
expected from the analysis of the single chain \cite{Lepri2011} (in that
work, the parameters were the same as used there). However, even a moderate
coupling, $\kappa =0.1$ (Fig.~\ref{f:fig5}b), restores the symmetry between
the chains, giving $t_{u}\approx t_{v}$.

\begin{figure}[th]
\begin{center}
\includegraphics[width=0.4\textwidth,clip]{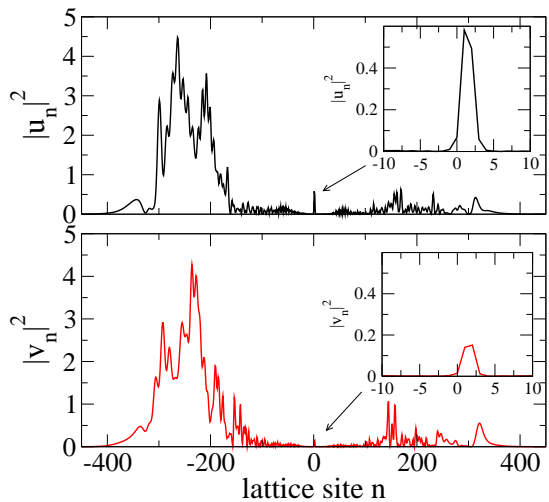}
\end{center}
\caption{(Color online) The snapshot of the norm densities in the two chains
at the end of the simulations, for the ladder with moderate coupling, $%
\protect\kappa =0.1$. The insets zoom the fields around the nonlinear sites,
showing that some energy remains trapped after the scattering. }
\label{f:snap}
\end{figure}

Figure \ref{f:snap} shows a snapshot of the norm densities at the end of the
same run which is presented in Fig.~\ref{f:fig5}b. Apart from the distortion
of the reflected and transmitted packets, it can been seen that some energy
remains trapped by the central segment in both cases, being larger in the
first one (see insets). A similar phenomenology was observed in the single
chain \cite{Lepri2012}, which is explained by the spontaneous formation of
localized defect modes at the nonlinear sites (see again the insets in Fig. %
\ref{f:snap}). The scattered packets remain almost monochromatic, at the
incident wave number $k_{0}$, with some weak radiation waves leaking
throughout the lattice.

\section{Symmetry breaking and restoration}

\label{sec:break}

The results of the previous Section indicate that the symmetry of the
scattered packets depends on the coupling. In the present Section we present
a more detailed analysis of the transition from symmetry-broken to symmetry
restored outputs as a function of the parameters. We anticipate that the
scenario may be considerably complex, as the transitions from the two
regimes may sensitively depend not only on the interchain coupling but also
on the nonlinearity level.

\subsection{The ladder-type system}

\begin{figure}[tph]
\begin{center}
\includegraphics[width=0.4\textwidth,clip ]{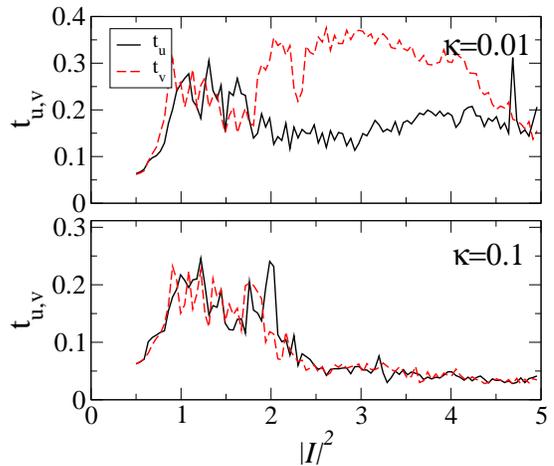}
\end{center}
\caption{(Color online) The dependence of the wavepacket transmission
coefficients $t_{u}$ (solid lines) and $t_{v}$ (dashed), see Eqs.~(\protect
\ref{tu}) and (\protect\ref{tv}), on the intensity $|I|^{2}$ for the ladder
model at two different interchain couplings $\protect\kappa $. Other
parameters are as in Fig. \protect\ref{f:fig5}.}
\label{fig:tco2}
\end{figure}

In Fig.~\ref{fig:tco2} we show the transmission coefficients, defined
according to Eqs.~(\ref{tu}) and (\ref{tv}), as functions of the
wavepacket's intensity, $|I|^{2}$. For the very weak coupling, there is a
region in which the two transmission coefficients are significantly
different (as in the example shown above), i.e., the diode-antidiode pair
still behaves similar to a single diode chain. However, the difference
between $t_{u}$ and $t_{v}$ tends to disappear with increasing $\kappa $.
For instance, at $\kappa =1$ the outputs are almost indistinguishable, even
at the central sites, which implies that the sufficiently strong coupling
makes the transmission effectively isotropic (bidirectional). The results
strongly depend on width $w$ of the incident packet, see Eq.~(\ref{gauss}).
For small $w$ the curves are much smoother, while broader packets (e.g.,
with $w=100$) give rise to wilder oscillations. Qualitatively speaking, this
can be understood by noting that the incoming wave packet acts as a
time-dependent force on the nonlinear core. Increasing $w$ amount to an
increase the power and thus the nonlinearity, resulting in more complex
(possibly chaotic) dynamics, which yields large changes in the transmission
and thus a sensitive dependence on the parameters of the incident packet.

\begin{figure}[tph]
\begin{center}
\includegraphics[width=0.4\textwidth,clip ]{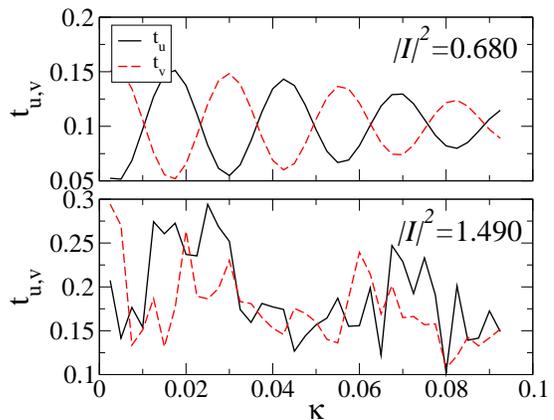}
\end{center}
\caption{(Color online) The dependence of the wavepacket transmission
coefficients $t_{u}$ (solid lines) and $t_{v}$ (dashed), see Eqs.~(\protect
\ref{tu}) and (\protect\ref{tv}), on coupling $\protect\kappa $ for the
ladder model at two different intensities. Other parameters are as in Fig.
\protect\ref{f:fig5}}
\label{fig:tco3}
\end{figure}

A noteworthy phenomenon is shown in Fig. \ref{fig:tco3}, where we display
the transmission coefficients as a function of the interchain coupling $%
\kappa $ for two fixed intensities, $|I|^{2}$. Namely, an antiphase
oscillating behavior of $t_{u}$ and $t_{v}$ is observed for the smaller
value of $|I|^{2}$.

\begin{figure}[tph]
\begin{center}
\includegraphics[width=0.4\textwidth,clip]{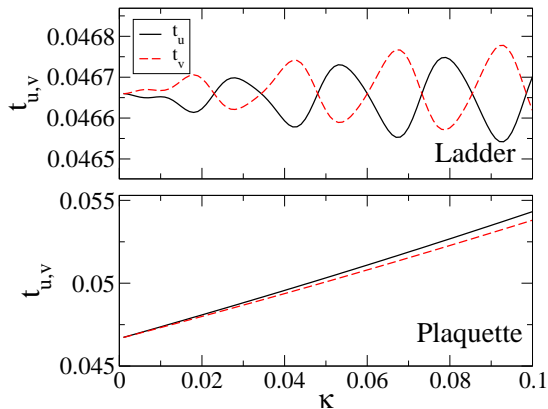}
\end{center}
\caption{(Color online) The dependence of the wavepacket transmission
coefficients $t_{u}$ (solid lines) and $t_{v}$ (dashed), see Eqs.~(\protect
\ref{tu}) and (\protect\ref{tv}), on coupling $\protect\kappa $ for $%
|I|^{2}=0.5$ in the linear system, with $\protect\alpha _{n}=\protect\beta %
_{n}=0$.}
\label{fig:linear}
\end{figure}

To understand the origin of such oscillations, in Fig. \ref{fig:linear} we report the limit case of the linear lattice, with 
$\alpha_{n}=\beta _{n}=0$  in Eq. (\ref{tdnls}). In this case, the same
oscillations occur too, although with a small amplitude. Thus, the
oscillations of the transmission coefficients in the linear ladder are
enhanced by a moderately strong nonlinearity. An essential point is that
this effect occurs in the ladder but not in the plaquette. This fact, as
well as the entire effect of the oscillations, is explained by the fact
that, in the ladder, the waves in the linearly coupled top and bottom chains
may periodically switch between them after the passage of the core segment
(the two central sites).

To analyze the oscillations in a quantitative form, we note that,
according to Eqs. (\ref{tdnls}), the temporal frequency of the energy
exchange between the chains is $\kappa $. In the combination with
the group velocity given by Eq. (\ref{V}), this implies that the spatial
period of the exchange in the wave packet propagating in the uniform linear
ladder is $L=2\pi V_{\mathrm{gr}}/\kappa \equiv 4\pi \kappa ^{-1}\sin k$. 
Thus, the intensity of the transmitted wave oscillates between the
chains, as a function of discrete coordinate $n$ as $\cos \left(
2\pi n/L\right) \equiv \cos \left( \left( 2\sin k\right) ^{-1}\kappa
n\right)$. At the edge of the lattice, $n=M$, this implies
oscillations between $t_{u}$ and $t_{v}$, following the
variation of $\kappa $, in the form of $\cos \left( \left( 2\sin
k\right) ^{-1}\kappa M\right) $. The period of the latter
oscillations is
\begin{equation}
\Delta \kappa =4\pi M^{-1}\sin k.  \label{Delta}
\end{equation}%
For the value $k=\pi /2$ fixed throughout the paper and for
$M=500$ (see caption to Fig. \ref{f:fig5}), Eq. (\ref{Delta}) yields
$\Delta \kappa =4\pi /500\approx 0.025$, which agrees with the
picture observed in Fig. \ref{fig:linear}.

To identify the symmetry-broken and symmetry-restored regions, we define the
asymmetry factor,
\begin{equation}
a=\frac{|t_{u}-t_{v}|}{t_{u}+t_{v}}  \label{a}
\end{equation}%
which is small when the output intensities in the two chains are nearly
equal, and approaches unity when one of the two transmission coefficients
vanishes.

\begin{figure}[tbp]
\begin{center}
\includegraphics[width=0.35\textwidth,clip ]{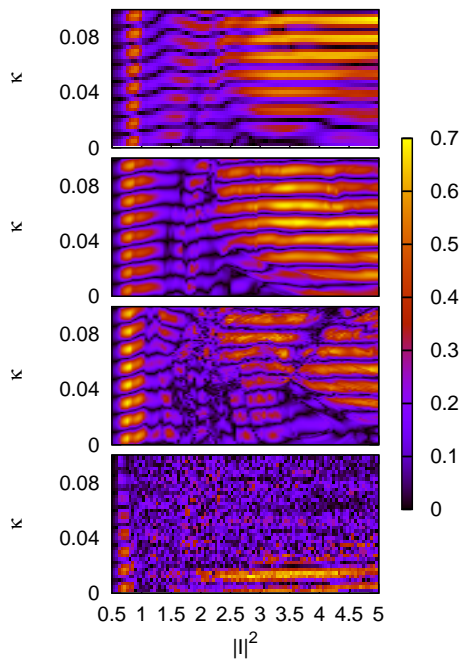}
\end{center}
\caption{(Color online) The ladder configuration with even initial
conditions, given by Eqs.~(\protect\ref{gauss}) with plus sign. Contour
plots of the asymmetry factor, (\protect\ref{a}), are shown for different
wavepacket's widths, $w=10,20,30,100$, from top to bottom. Other parameters
are as in Fig. \protect\ref{f:fig5}.}
\label{fig:as}
\end{figure}

\begin{figure}[tbp]
\begin{center}
\includegraphics[width=0.35\textwidth,clip ]{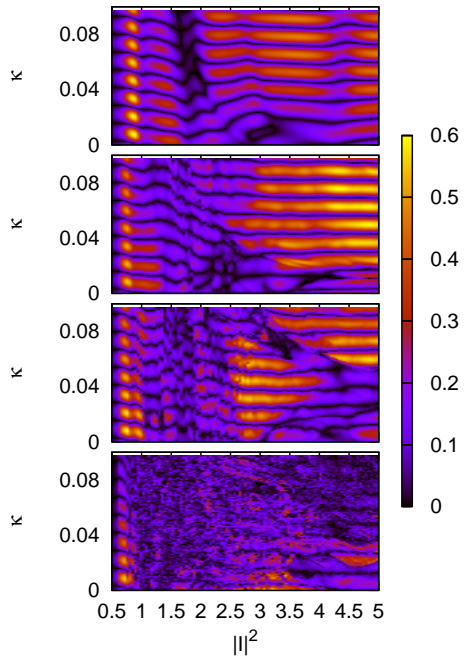}
\end{center}
\caption{(Color online) The ladder configuration with odd initial
conditions, given by Eqs.~(\protect\ref{gauss}) with the minus sign. Contour
plots of the asymmetry factor, (\protect\ref{a}), are shown for different
wavepacket's widths, $w=10,20,30,100$, from top to bottom. Other parameters
are as in Fig. \protect\ref{f:fig5}. }
\label{fig:asa}
\end{figure}

In Figs. \ref{fig:as} and \ref{fig:asa} we show contour plots of $a$ in the $%
(|I|^{2},\kappa )$ plane for different wavepacket widths $w$ in the ladder
model with even and odd initial conditions, respectively, in Eqs.~(\ref%
{gauss}). It is seen that there are parameter ranges where $a$ is very
small, alternating with many regions where the symmetry breaking is
sizeable. Roughly, the regions of small $a$ are concentrated around $%
|I|^{2}\approx 0.6$ (in the leftmost part of the figures) and for larger
intensities and couplings---around $|I|^{2}\gtrsim 2.5$, $\kappa \gtrsim
0.02 $. Notably, there are oscillations caused by the antiphase behavior
observed in Fig \ref{fig:tco3}.

%\begin{figure}
% \begin{center}
% \includegraphics[width=1.4\textwidth,clip ]{comparet_all.eps}
% \caption{Ladder case, even initial conditions: Contour plots of transmission coefficients
% for different packet widths, $w=10, 20, 30, 100$ from top to bottom.
% Other parameters as in fig. \ref{f:fig5}}
% \label{fig:ct}
%\end{center}
%\end{figure}

%\begin{figure}
% \begin{center}
% \includegraphics[width=1.4\textwidth,clip ]{comparet_asm.eps}
% \caption{Ladder case, odd initial conditions:
% Contour plots of transmission coefficients
% for different packet widths, $w=10, 20, 30, 100$ from top to bottom.
% Other parameters as in fig. \ref{f:fig5}}
% \label{fig:cta}
%\end{center}
%\end{figure}

%\begin{figure}
% \begin{center}
% \includegraphics[width=1.4\textwidth,clip ]{comparet_plq.eps}
% \caption{Plaquette case, even initial conditions: Contour plots of transmission coefficients
% for different packet widths, $w=10, 20, 30, 100$ from top to bottom.
% Other parameters as in fig. \ref{f:fig5}}
% \label{fig:ctp}
%\end{center}
%\end{figure}

\subsection{The plaquette-type system}

\begin{figure}[th]
\begin{center}
\includegraphics[width=0.35\textwidth,clip ]{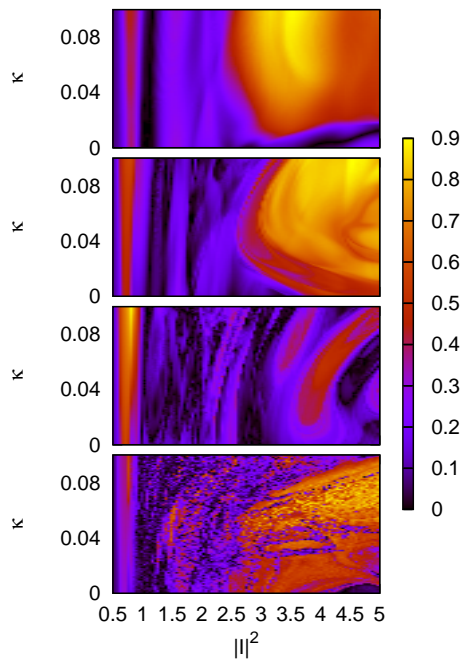}
\end{center}
\caption{(Color online) The plaquette configuration, with the even initial
conditions, given by Eqs.~(\protect\ref{gauss}) with the plus sign. Contour
plots of the asymmetry factor, (\protect\ref{a}), are shown for different
wavepacket's widths, $w=10,20,30,100$ from top to bottom. Other parameters
are as in Fig. \protect\ref{f:fig5}}
\label{fig:asp}
\end{figure}

As seen in Fig.~\ref{fig:asp}, the structure of symmetry-broken regions is
much simpler for the plaquette configuration. This is explained by the fact
that the absence of the linear coupling between the chains outside of the
central segment does not allow more subtle feature to develop via the
continuing switching of the power between the chains.

\begin{figure}[tbp]
\begin{center}
\includegraphics[width=0.35\textwidth,clip]{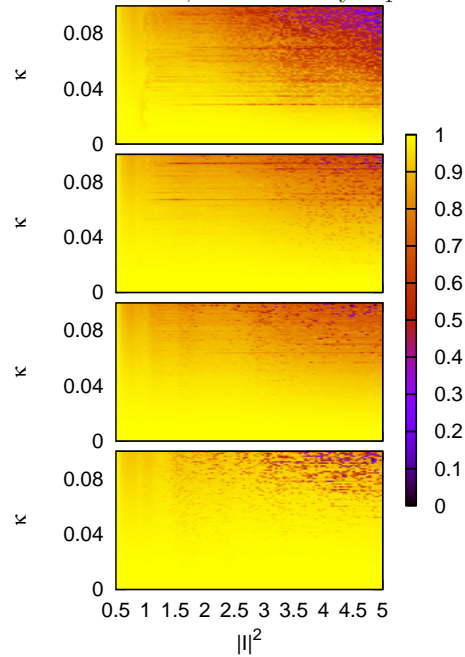}
\end{center}
\caption{(Color online) The plaquette configuration with initial conditions
Eq.~(\protect\ref{asymu}) corresponding to the wavepacket initially located
in the single chain. Contour plots of the asymmetry factor, (\protect\ref{a}%
), are displayed for different wavepacket's widths, $w=10,20,30,100$ from
top to bottom. Other parameters as in Fig. \protect\ref{f:fig5}.}
\label{fig:aspu}
\end{figure}

In the case of the plaquette, it is also possible to launch an input in a
single chain, as the chains are decoupled outside of the central segment. In
this vein, we have investigated the asymmetric initial conditions with the
wavepacket initially launched in one of the two chains:
\begin{eqnarray}
&&u_{n}(0)\;=\;I\exp \left[ -\frac{(n-n_{0})^{2}}{w^{2}}+ik_{0}n\right] ,
\notag \\
&&v_{n}(0)\;=\;0,  \label{asymu}
\end{eqnarray}%
as well as
\begin{eqnarray}
&&u_{n}(0)\;=\;0,  \notag \\
&&v_{n}(0)\;=\;I\exp \left[ -\frac{(n-n_{0})^{2}}{w^{2}}+ik_{0}n\right] .
\label{asymv}
\end{eqnarray}%
For comparison with the previous cases, we consider the same ranges of
coupling parameters and intensities. Here, the results are rather simple, as
shown in Fig. \ref{fig:aspu}: The growth of the intensity, $I$, i.e., of the
nonlinearity strength, causes the wave to stay in the original chain, while
the increase of the coupling constant, $\kappa $, pushes the system towards
even distribution of the intensity between the coupled chains.

%in Figs. \ref{fig:ctpu}  \ref{fig:ctpv}  \ref{fig:aspu} \ref{fig:aspv}.

\section{Conclusions}

\label{sec:fin}

We have presented numerical simulations for the scattering of wavepackets in
a model of two parallel-coupled lattices, each of them acting individually
as diode chains in opposite directions. We discussed two different
configurations, the ladder and the plaquette. In the case of weak linear
coupling, the symmetry breaking is observed (i.e., the pair still features
the diode behavior), while the moderately strong coupling gradually restores
the symmetry (makes the transmission effectively bidirectional). In the case
of the ladder, an oscillatory dependence of the transmission on the strength
of the coupling is observed. Such oscillations are qualitatively explained
as nonlinear versions of those occurring in the linear limit. As a result,
the symmetry-broken regions in the parameter space form a complex pattern,
alternating with regions of the restored symmetry. The location of such
regions are also strongly dependent on the wavepacket widths.

As mentioned in the Introduction, one our motivations was to study
asymmetric wavepacket propagation \cite{Lepri2011}. In this context,
the present study is a first attempt to discuss a ``multichannel"
configuration in which several inputs can be simultaneously managed
through nonlinear interactions. We outlined how the concepts of
symmetry breaking and restoration are important issues. Indeed, they
directly relate to the functionality of the proposed system as a
tool to selectively control the transmission direction of multiple
inputs. Moreover, the present paper is the a starting point to
address reciprocity violations in general lattices with nontrivial
topologies and/or couplings. This work can be extended by
considering more complex setups where multiple ``leads" are coupled
through nonlinear sites. In particular, it may be interesting to
consider a network built of alternating parallel-coupled chains with
opposite transmission directions.

%\acknowledgements
%
% \bibliography{diodo}{}

\begin{thebibliography}{31}
\expandafter\ifx\csname natexlab\endcsname\relax\def\natexlab#1{#1}\fi
\expandafter\ifx\csname bibnamefont\endcsname\relax
  \def\bibnamefont#1{#1}\fi
\expandafter\ifx\csname bibfnamefont\endcsname\relax
  \def\bibfnamefont#1{#1}\fi
\expandafter\ifx\csname citenamefont\endcsname\relax
  \def\citenamefont#1{#1}\fi
\expandafter\ifx\csname url\endcsname\relax
  \def\url#1{\texttt{#1}}\fi
\expandafter\ifx\csname urlprefix\endcsname\relax\def\urlprefix{URL }\fi
\providecommand{\bibinfo}[2]{#2}
\providecommand{\eprint}[2][]{\url{#2}}

\bibitem[{\citenamefont{Kosevich}(1995)}]{Kosevich1995}
\bibinfo{author}{\bibfnamefont{Y.~A.} \bibnamefont{Kosevich}},
  \bibinfo{journal}{Phys. Rev. B} \textbf{\bibinfo{volume}{52}},
  \bibinfo{pages}{1017} (\bibinfo{year}{1995}).

\bibitem[{\citenamefont{Liang et~al.}(2009)\citenamefont{Liang, Yuan, and chun
  Cheng}}]{Liang09}
\bibinfo{author}{\bibfnamefont{B.}~\bibnamefont{Liang}},
  \bibinfo{author}{\bibfnamefont{B.}~\bibnamefont{Yuan}}, \bibnamefont{and}
  \bibinfo{author}{\bibfnamefont{J.}~\bibnamefont{chun Cheng}},
  \bibinfo{journal}{Phys. Rev. Lett.} \textbf{\bibinfo{volume}{103}},
  \bibinfo{eid}{104301} (\bibinfo{year}{2009}).

\bibitem[{\citenamefont{Liang et~al.}(2010)\citenamefont{Liang, Guo, Tu, Zhang,
  and Cheng}}]{Liang2010}
\bibinfo{author}{\bibfnamefont{B.}~\bibnamefont{Liang}},
  \bibinfo{author}{\bibfnamefont{X.}~\bibnamefont{Guo}},
  \bibinfo{author}{\bibfnamefont{J.}~\bibnamefont{Tu}},
  \bibinfo{author}{\bibfnamefont{D.}~\bibnamefont{Zhang}}, \bibnamefont{and}
  \bibinfo{author}{\bibfnamefont{J.}~\bibnamefont{Cheng}},
  \bibinfo{journal}{Nature Materials} \textbf{\bibinfo{volume}{9}},
  \bibinfo{pages}{989} (\bibinfo{year}{2010}), ISSN \bibinfo{issn}{1476-1122}.

\bibitem[{\citenamefont{Nesterenko et~al.}(2005)\citenamefont{Nesterenko,
  Daraio, Herbold, and Jin}}]{Nesterenko05}
\bibinfo{author}{\bibfnamefont{V.~F.} \bibnamefont{Nesterenko}},
  \bibinfo{author}{\bibfnamefont{C.}~\bibnamefont{Daraio}},
  \bibinfo{author}{\bibfnamefont{E.~B.} \bibnamefont{Herbold}},
  \bibnamefont{and} \bibinfo{author}{\bibfnamefont{S.}~\bibnamefont{Jin}},
  \bibinfo{journal}{Phys. Rev. Lett.} \textbf{\bibinfo{volume}{95}},
  \bibinfo{pages}{1} (\bibinfo{year}{2005}).

\bibitem[{\citenamefont{Boechler et~al.}(2011)\citenamefont{Boechler,
  Theocharis, and Daraio}}]{Boechler2011}
\bibinfo{author}{\bibfnamefont{N.}~\bibnamefont{Boechler}},
  \bibinfo{author}{\bibfnamefont{G.}~\bibnamefont{Theocharis}},
  \bibnamefont{and} \bibinfo{author}{\bibfnamefont{C.}~\bibnamefont{Daraio}},
  \bibinfo{journal}{Nature Materials} \textbf{\bibinfo{volume}{10}},
  \bibinfo{pages}{665} (\bibinfo{year}{2011}).

\bibitem[{\citenamefont{Scalora et~al.}(1994)\citenamefont{Scalora, Dowling,
  Bowden, and Bloemer}}]{Scalora94}
\bibinfo{author}{\bibfnamefont{M.}~\bibnamefont{Scalora}},
  \bibinfo{author}{\bibfnamefont{J.~P.} \bibnamefont{Dowling}},
  \bibinfo{author}{\bibfnamefont{C.~M.} \bibnamefont{Bowden}},
  \bibnamefont{and} \bibinfo{author}{\bibfnamefont{M.~J.}
  \bibnamefont{Bloemer}}, \bibinfo{journal}{J. Appl. Phys.}
  \textbf{\bibinfo{volume}{76}}, \bibinfo{pages}{2023} (\bibinfo{year}{1994}).

\bibitem[{\citenamefont{Tocci et~al.}(1995)\citenamefont{Tocci, Bloemer,
  Scalora, Dowling, and Bowden}}]{Tocci95}
\bibinfo{author}{\bibfnamefont{M.~D.} \bibnamefont{Tocci}},
  \bibinfo{author}{\bibfnamefont{M.~J.} \bibnamefont{Bloemer}},
  \bibinfo{author}{\bibfnamefont{M.}~\bibnamefont{Scalora}},
  \bibinfo{author}{\bibfnamefont{J.~P.} \bibnamefont{Dowling}},
  \bibnamefont{and} \bibinfo{author}{\bibfnamefont{C.~M.}
  \bibnamefont{Bowden}}, \bibinfo{journal}{Appl. Phys. Lett.}
  \textbf{\bibinfo{volume}{66}}, \bibinfo{pages}{2324} (\bibinfo{year}{1995}).

\bibitem[{\citenamefont{Gallo et~al.}(2001)\citenamefont{Gallo, Assanto,
  Parameswaran, and Fejer}}]{Gallo01}
\bibinfo{author}{\bibfnamefont{K.}~\bibnamefont{Gallo}},
  \bibinfo{author}{\bibfnamefont{G.}~\bibnamefont{Assanto}},
  \bibinfo{author}{\bibfnamefont{K.}~\bibnamefont{Parameswaran}},
  \bibnamefont{and} \bibinfo{author}{\bibfnamefont{M.}~\bibnamefont{Fejer}},
  \bibinfo{journal}{Appl. Phys. Lett.} \textbf{\bibinfo{volume}{79}},
  \bibinfo{pages}{314} (\bibinfo{year}{2001}).

\bibitem[{\citenamefont{Feise et~al.}(2005)\citenamefont{Feise, Shadrivov, and
  Kivshar}}]{Feise05}
\bibinfo{author}{\bibfnamefont{M.~W.} \bibnamefont{Feise}},
  \bibinfo{author}{\bibfnamefont{I.~V.} \bibnamefont{Shadrivov}},
  \bibnamefont{and} \bibinfo{author}{\bibfnamefont{Y.~S.}
  \bibnamefont{Kivshar}}, \bibinfo{journal}{Phys. Rev. E}
  \textbf{\bibinfo{volume}{71}}, \bibinfo{eid}{037602} (\bibinfo{year}{2005}).

\bibitem[{\citenamefont{Konotop and Kuzmiak}(2002)}]{Konotop02}
\bibinfo{author}{\bibfnamefont{V.~V.} \bibnamefont{Konotop}} \bibnamefont{and}
  \bibinfo{author}{\bibfnamefont{V.}~\bibnamefont{Kuzmiak}},
  \bibinfo{journal}{Phys. Rev. B} \textbf{\bibinfo{volume}{66}},
  \bibinfo{eid}{235208} (\bibinfo{year}{2002}).

\bibitem[{\citenamefont{Biancalana}(2008)}]{Biancalana08}
\bibinfo{author}{\bibfnamefont{F.}~\bibnamefont{Biancalana}},
  \bibinfo{journal}{J. Appl. Phys.} \textbf{\bibinfo{volume}{104}},
  \bibinfo{pages}{093113} (\bibinfo{year}{2008}).

\bibitem[{\citenamefont{Grigoriev and Biancalana}(2011)}]{Grigoriev2011}
\bibinfo{author}{\bibfnamefont{V.}~\bibnamefont{Grigoriev}} \bibnamefont{and}
  \bibinfo{author}{\bibfnamefont{F.}~\bibnamefont{Biancalana}},
  \bibinfo{journal}{Opt. Lett.} \textbf{\bibinfo{volume}{36}},
  \bibinfo{pages}{2131} (\bibinfo{year}{2011}).

\bibitem[{\citenamefont{Ramezani et~al.}(2010)\citenamefont{Ramezani, Kottos,
  El-Ganainy, and Christodoulides}}]{Ramezani2010}
\bibinfo{author}{\bibfnamefont{H.}~\bibnamefont{Ramezani}},
  \bibinfo{author}{\bibfnamefont{T.}~\bibnamefont{Kottos}},
  \bibinfo{author}{\bibfnamefont{R.}~\bibnamefont{El-Ganainy}},
  \bibnamefont{and} \bibinfo{author}{\bibfnamefont{D.~N.}
  \bibnamefont{Christodoulides}}, \bibinfo{journal}{Phys. Rev. A}
  \textbf{\bibinfo{volume}{82}}, \bibinfo{pages}{043803}
  (\bibinfo{year}{2010}).

\bibitem[{\citenamefont{D'Ambroise et~al.}(2012)\citenamefont{D'Ambroise,
  Kevrekidis, and Lepri}}]{D'Ambroise2012}
\bibinfo{author}{\bibfnamefont{J.}~\bibnamefont{D'Ambroise}},
  \bibinfo{author}{\bibfnamefont{P.~G.} \bibnamefont{Kevrekidis}},
  \bibnamefont{and} \bibinfo{author}{\bibfnamefont{S.}~\bibnamefont{Lepri}},
  \bibinfo{journal}{J. Phys. A: Math. Theor.} \textbf{\bibinfo{volume}{45}},
  \bibinfo{pages}{444012} (\bibinfo{year}{2012}).

\bibitem[{\citenamefont{Roy}(2010)}]{Roy2010}
\bibinfo{author}{\bibfnamefont{D.}~\bibnamefont{Roy}}, \bibinfo{journal}{Phys.
  Rev. B} \textbf{\bibinfo{volume}{81}}, \bibinfo{pages}{155117}
  (\bibinfo{year}{2010}).

\bibitem[{\citenamefont{Tao et~al.}(2011)\citenamefont{Tao, Chen, Xu, Pan, and
  Du}}]{Tao2011}
\bibinfo{author}{\bibfnamefont{F.}~\bibnamefont{Tao}},
  \bibinfo{author}{\bibfnamefont{W.}~\bibnamefont{Chen}},
  \bibinfo{author}{\bibfnamefont{W.}~\bibnamefont{Xu}},
  \bibinfo{author}{\bibfnamefont{J.}~\bibnamefont{Pan}}, \bibnamefont{and}
  \bibinfo{author}{\bibfnamefont{S.}~\bibnamefont{Du}}, \bibinfo{journal}{Phys.
  Rev. E} \textbf{\bibinfo{volume}{83}}, \bibinfo{pages}{056605}
  (\bibinfo{year}{2011}).

\bibitem[{\citenamefont{Narayan and Dhar}(2004)}]{Narayan04}
\bibinfo{author}{\bibfnamefont{O.}~\bibnamefont{Narayan}} \bibnamefont{and}
  \bibinfo{author}{\bibfnamefont{A.}~\bibnamefont{Dhar}},
  \bibinfo{journal}{Europhys. Lett.} \textbf{\bibinfo{volume}{67}},
  \bibinfo{pages}{559} (\bibinfo{year}{2004}).

\bibitem[{\citenamefont{Maznev et~al.}(2012)\citenamefont{Maznev, Every, and
  Wright}}]{Maznev2012}
\bibinfo{author}{\bibfnamefont{A.}~\bibnamefont{Maznev}},
  \bibinfo{author}{\bibfnamefont{A.}~\bibnamefont{Every}}, \bibnamefont{and}
  \bibinfo{author}{\bibfnamefont{O.}~\bibnamefont{Wright}},
  \bibinfo{journal}{arXiv preprint arXiv:1210.6309}  (\bibinfo{year}{2012}).

\bibitem[{\citenamefont{Eilbeck et~al.}(1985)\citenamefont{Eilbeck, Lomdahl,
  and Scott}}]{Eilbeck1985}
\bibinfo{author}{\bibfnamefont{J.~C.} \bibnamefont{Eilbeck}},
  \bibinfo{author}{\bibfnamefont{P.~S.} \bibnamefont{Lomdahl}},
  \bibnamefont{and} \bibinfo{author}{\bibfnamefont{A.~C.} \bibnamefont{Scott}},
  \bibinfo{journal}{Physica D} \textbf{\bibinfo{volume}{16}},
  \bibinfo{pages}{318} (\bibinfo{year}{1985}).

\bibitem[{\citenamefont{Brazhnyi and Malomed}(2011)}]{Brazhnyi2011}
\bibinfo{author}{\bibfnamefont{V.}~\bibnamefont{Brazhnyi}} \bibnamefont{and}
  \bibinfo{author}{\bibfnamefont{B.~A.} \bibnamefont{Malomed}},
  \bibinfo{journal}{Phys. Rev. A} \textbf{\bibinfo{volume}{83}},
  \bibinfo{pages}{053844} (\bibinfo{year}{2011}).

\bibitem[{\citenamefont{Molina and Tsironis}(1993)}]{Molina1993}
\bibinfo{author}{\bibfnamefont{M.~I.} \bibnamefont{Molina}} \bibnamefont{and}
  \bibinfo{author}{\bibfnamefont{G.~P.} \bibnamefont{Tsironis}},
  \bibinfo{journal}{Phys. Rev. B} \textbf{\bibinfo{volume}{47}},
  \bibinfo{pages}{15330} (\bibinfo{year}{1993}).

\bibitem[{\citenamefont{Gupta and Kundu}(1997{\natexlab{a}})}]{Gupta1997}
\bibinfo{author}{\bibfnamefont{B.~C.} \bibnamefont{Gupta}} \bibnamefont{and}
  \bibinfo{author}{\bibfnamefont{K.}~\bibnamefont{Kundu}},
  \bibinfo{journal}{Phys. Rev. B} \textbf{\bibinfo{volume}{55}},
  \bibinfo{pages}{894} (\bibinfo{year}{1997}{\natexlab{a}}).

\bibitem[{\citenamefont{Gupta and Kundu}(1997{\natexlab{b}})}]{Gupta1997a}
\bibinfo{author}{\bibfnamefont{B.~C.} \bibnamefont{Gupta}} \bibnamefont{and}
  \bibinfo{author}{\bibfnamefont{K.}~\bibnamefont{Kundu}},
  \bibinfo{journal}{Phys. Rev. B} \textbf{\bibinfo{volume}{55}},
  \bibinfo{pages}{11033} (\bibinfo{year}{1997}{\natexlab{b}}).

\bibitem[{\citenamefont{Bulgakov et~al.}(2011)\citenamefont{Bulgakov, Pichugin,
  and Sadreev}}]{Bulgakov2011}
\bibinfo{author}{\bibfnamefont{E.}~\bibnamefont{Bulgakov}},
  \bibinfo{author}{\bibfnamefont{K.}~\bibnamefont{Pichugin}}, \bibnamefont{and}
  \bibinfo{author}{\bibfnamefont{A.}~\bibnamefont{Sadreev}},
  \bibinfo{journal}{Phys. Rev. B} \textbf{\bibinfo{volume}{83}},
  \bibinfo{pages}{045109} (\bibinfo{year}{2011}).

\bibitem[{\citenamefont{Kevrekidis}(2009)}]{Kevrekidis}
\bibinfo{author}{\bibfnamefont{P.~G.} \bibnamefont{Kevrekidis}},
  \emph{\bibinfo{title}{The Discrete Nonlinear Schr\"odinger Equation}}
  (\bibinfo{publisher}{Springer Verlag, Berlin}, \bibinfo{year}{2009}).

\bibitem[{\citenamefont{Eilbeck and Johansson}(2003)}]{Eilbeck2003}
\bibinfo{author}{\bibfnamefont{J.~C.} \bibnamefont{Eilbeck}} \bibnamefont{and}
  \bibinfo{author}{\bibfnamefont{M.}~\bibnamefont{Johansson}}, in
  \emph{\bibinfo{booktitle}{Conference on Localization and Energy Transfer in
  Nonlinear Systems}}, edited by
  \bibinfo{editor}{\bibfnamefont{L.}~\bibnamefont{Vazquez}},
  \bibinfo{editor}{\bibfnamefont{R.~S.} \bibnamefont{MacKay}},
  \bibnamefont{and} \bibinfo{editor}{\bibfnamefont{M.~P.}
  \bibnamefont{Zorzano}} (\bibinfo{publisher}{World Scientific, Singapore},
  \bibinfo{year}{2003}), p.~\bibinfo{pages}{44}.

\bibitem[{\citenamefont{Lepri and Casati}(2011)}]{Lepri2011}
\bibinfo{author}{\bibfnamefont{S.}~\bibnamefont{Lepri}} \bibnamefont{and}
  \bibinfo{author}{\bibfnamefont{G.}~\bibnamefont{Casati}},
  \bibinfo{journal}{Phys. Rev. Lett.} \textbf{\bibinfo{volume}{106}},
  \bibinfo{pages}{164101} (\bibinfo{year}{2011}).

\bibitem[{\citenamefont{Herring et~al.}(2007)\citenamefont{Herring, Kevrekidis,
  Malomed, Carretero-Gonz{\'a}lez, and Frantzeskakis}}]{Herring2007}
\bibinfo{author}{\bibfnamefont{G.}~\bibnamefont{Herring}},
  \bibinfo{author}{\bibfnamefont{P.~G.} \bibnamefont{Kevrekidis}},
  \bibinfo{author}{\bibfnamefont{B.~A.} \bibnamefont{Malomed}},
  \bibinfo{author}{\bibfnamefont{R.}~\bibnamefont{Carretero-Gonz{\'a}lez}},
  \bibnamefont{and}
  \bibinfo{author}{\bibfnamefont{D.}~\bibnamefont{Frantzeskakis}},
  \bibinfo{journal}{Phys. Rev. E} \textbf{\bibinfo{volume}{76}},
  \bibinfo{pages}{066606} (\bibinfo{year}{2007}).

\bibitem[{\citenamefont{Malomed}(2013)}]{Malomed2013}
\bibinfo{editor}{\bibfnamefont{B.~A.} \bibnamefont{Malomed}}, ed.,
  \emph{\bibinfo{title}{Spontaneous Symmetry Breaking, Self-trapping, and
  Josephson Oscillations}} (\bibinfo{publisher}{Springer-Verlag Berlin and
  Heidelberg}, \bibinfo{year}{2013}).

\bibitem[{\citenamefont{Had{\v{z}}ievski
  et~al.}(2010)\citenamefont{Had{\v{z}}ievski, Gligori{\'c}, Maluckov, and
  Malomed}}]{Hadvzievski2010}
\bibinfo{author}{\bibfnamefont{L.}~\bibnamefont{Had{\v{z}}ievski}},
  \bibinfo{author}{\bibfnamefont{G.}~\bibnamefont{Gligori{\'c}}},
  \bibinfo{author}{\bibfnamefont{A.}~\bibnamefont{Maluckov}}, \bibnamefont{and}
  \bibinfo{author}{\bibfnamefont{B.~A.} \bibnamefont{Malomed}},
  \bibinfo{journal}{Phys. Rev. A} \textbf{\bibinfo{volume}{82}},
  \bibinfo{pages}{033806} (\bibinfo{year}{2010}).

\bibitem[{\citenamefont{Lepri and Casati}(2012)}]{Lepri2012}
\bibinfo{author}{\bibfnamefont{S.}~\bibnamefont{Lepri}} \bibnamefont{and}
  \bibinfo{author}{\bibfnamefont{G.}~\bibnamefont{Casati}},
  \bibinfo{journal}{arXiv preprint arXiv:1211.4996}  (\bibinfo{year}{2012}).

\end{thebibliography}
% \bibliographystyle{apsrev}

\end{document}